\documentstyle[12pt]{article}
\normalbaselines
\begin{document}

\title{Evolution Kernels of Twist-3 Light-Ray Operators 
in Polarized Deep Inelastic Scattering}
\author{B. Geyer, D. M\"uller, D. Robaschik \\
\hspace{0.01cm}
 Fakult\"at f\"ur Physik und Geowissenschaft der Universit\"at Leipzig\\
         Augustusplatz, D-04109 Leipzig, Germany}

\date{\ }
\maketitle
 
\thispagestyle{empty}

\begin{abstract}

The twist three contributions to the $Q^2$-evolution of the spin-dependent
structure function $g_2(x)$ are considered in the non-local operator
product approach. Starting from the perturbative expansion of the T-product
of two electromagnetic currents, we introduce the nonlocal light-cone expansion
proved by Anikin and Zavialov  and  determine  the physical relevant set of
light-ray operators of twist three. Using the equations of motion we show the
equivalence of these operators to the Shuryak-Vainshtein operators plus the
mass operator, and we determine their evolution kernels using the light-cone
gauge with the Leibbrandt-Mandelstam prescription. The result of Balitsky and
Braun for the twist three evolution kernel (nonsinglet case) is confirmed.

\end{abstract}

\section{Introduction}
\setcounter{equation}{0}
\renewcommand{\theequation}{\arabic{section}.\arabic{equation}}

According  to  the  present  status  of  quantum  chromodynamics 
nonperturbative inputs, being not calculable from first principles,
are  necessary  for  the interpretation and prediction of experimental
results.  Nevertheless, these inputs, for instance parton distribution
and   fragmentation   functions,  satisfy  evolution  equations  whose
integral  kernels  are  calculable  perturbatively.  For the twist two
parton distribution functions all evolution kernels are now known
up to next-to-leading order.

However, also higher twist effects are experimentally accessible. 
Recently, in deep inelastic scattering the first moments of the polarized 
structure function  $g_2(x)$ 
are measured \cite{g2}. In leading order of the momentum transfer
$Q^2$ this structure function is determined  by twist two 
as well as twist three contributions.
In  comparison  with  the  twist two case the treatment of twist three
 is  technically  more  subtle  due to the appearance of a set of
operators  mixing  under  renormalization and constrained by relations
between themselves. Up to now there exist already several papers which
determine  the  local anomalous dimensions \cite{BKL,JI,KO} as well as
nonlocal operators and the evolution kernels for the distribution functions
\cite{BKL,RAT,BB}.

The  two most complete calculations of the evolution kernels including
the  flavor  singlet  case,  \cite{BKL}  and \cite{BB}, are based on 
complicated   techniques;   moreover,   the   comparison  between  both
approaches  is  by  no means trivial and both papers contain also some
misprints. Therefore, in a first step, our aim is to confirm the known
results  using  an  independent approach and compare it with the local
anomalous dimensions. In a next step we will study the solution of the
evolution  equations  including  the singlet case. Here we present the
first  results  for  the  nonsinglet  case.  Technically  we apply the
nonlocal  operator  product expansion \cite{AZ}, \cite{FORT} proved by
Anikin and Zavialov in renormalized quantum field theory. To avoid the
complications with ghost and gauge-variant operators which are present
in  covariant gauge we apply the light-cone gauge using the 
Leibbrand-Mandelstam prescription \cite{LM}, \cite{BAS}.

\section{Perturbative Analysis}
\setcounter{equation}{0}
\renewcommand{\theequation}{\arabic{section}.\arabic{equation}}

The  part  of  the  hadronic  tensor being relevant for polarized deep
inelastic  scattering  is  determined  by the antisymmetric part (with
respect  to  Lorentz  indices)  of  the absorptive part of the virtual
forward  Compton amplitude. In the Bjorken region the leading terms in
$Q^2$  of  the  Compton amplitude correspond to the leading light-cone
singularities in the coordinate space.

Our  analysis  starts  with  a  perturbative investigation of the time
ordered  product  of  two electromagnetic currents. For simplicity, we
will  neglect  flavor  and  color indices in the following. In leading
order the antisymmetric part is determined by both twist two and twist
three contributions. A perturbative expansion up to the first order in
the   coupling   constant  $ g$   provides   additionally   twist   four
contributions:
\begin{eqnarray}
\label{p1.4}
\lefteqn{
 \{T{\bar\psi}(x)\gamma^\mu \psi(x)
   {\bar\psi}(y)\gamma^\nu \psi(y)\}^{\rm as} =
    \epsilon^{\mu\nu\rho\sigma} \{i\partial_\rho^{x} D^c(x,y)\;
    {\bar\psi}(x)\gamma_\sigma\gamma_5 U(x,y)
    \psi(y) } \\
 & & +\;{g\over 2} D^c(x,y)\; {\bar\psi}(x)\gamma_\sigma\gamma_5  
    \int_0^1 dt \left(2 t- 1\right)U(x,z)
F_{\rho\lambda}(z) (x-y)^\lambda
     U(z,y) \psi(y)\nonumber \\
 & & -\;{ig\over 2} D^c(x,y)\;
 {\bar\psi}(x)\gamma_\sigma  
    \int_0^1 dt U(x,z)  {\tilde F}_{\rho\lambda}(z)
    (x-y)^\lambda
     U(z,y) \psi(y) \}
     + (x,\mu  \leftrightarrow y, \nu), \nonumber
\end{eqnarray}
where $z=z(t)$ with $z(1)=x$ and $z(0)=y$;
${\tilde F}_{\alpha\beta}= 1/2 \epsilon_{\alpha\beta\mu\nu} F^{\mu\nu}$ is the
dual field strength tensor and the path ordered phase factor
\begin{eqnarray}
\label{p1.5 }
      U(x,y)& =& P \exp\left\{-ig\int_y^x dw^\mu  A_{\mu}(w)\right\}
\end{eqnarray}
ensures gauge invariance.
Instead of the free  field  propagator $S_0^c(x)$
we used here the spinor propagator in
an external gluon field (which is a straightforward generalization of the
abelian case \cite{GT}). One gets up to the first order in $g$: 
\begin{eqnarray*}
 S^c(x,0)&=& S^c_0(x)\; U_s(x,0) + \\
& &  {g\over 4} D^c(x) \int_0^1 dt
\left\{(4t-2)\gamma^\mu x^\nu - \gamma^\mu\not\! x 
      \gamma^\nu \right\} U_s(x,tx) F_{\mu \nu}(tx) U_s(tx,0) +\cdots, \\
 \mbox{with}\quad\quad & & \nonumber\\
U_s(x,y)   &=& P \exp\left\{-ig \int_0^1 d\lambda A_\mu(w(\lambda))
(x^\mu-y^\mu)\right\},
 \quad  w(\lambda) = x \lambda + y (1-\lambda).
\end{eqnarray*}

The  nonlocal  light-cone expansion is obtained by approximating 
the  vector  $x$  by  the  light-like  vector  $\tilde  x$  defined as
$x=\tilde{x} + a(x,\eta) \eta $, where $\eta$ denotes a fixed auxiliary vector
(e.g. normalized by  $\eta^2  = 1$) and $a$ the corresponding coefficient. In
leading order we substitute $px \rightarrow p\tilde{x} $, where $p$ denotes
a   momentum,   but  the  $x^2$-singularities  remain  unchanged,  $x^2
\rightarrow  x^2$. In general, the coefficient function will depend 
on two auxiliary variables $\kappa_i$, whose range (according to the
$\alpha$-representation of the contributing Feynman diagrams) is restricted by  
$0\leq \kappa_i \leq 1$.  Here we introduce these variables quite
 trivially through integration over two $\delta$-functions. 
In this more or less intuitive way we get the following light-cone expansion:
\begin{eqnarray}
\label{lc1}
\lefteqn{
\hspace{-1.7cm}  \{Tj^\mu(x) j^\nu(0)\}^{\rm as}  = 
 \epsilon^{\mu\nu\rho\sigma}
 \int^1_0 d{\kappa_1}\int^1_0 d{\kappa_2}
  \delta(\kappa_1 -1) \delta(\kappa_2 ) } \nonumber \\
& & \quad\quad\Big\{i\partial_\rho^{x} D^c(x)\;
    {\bar\psi}(\kappa_1 \tilde x)\gamma_\sigma\gamma_5 U_s(\kappa_1
    \tilde x,\kappa_2 \tilde x)   \psi(\kappa_2 \tilde x)  \\
& & \quad\quad \; -\; {g\over 2} D^c(x)\;
    {\bar\psi}(\kappa_1 \tilde x)\gamma_\sigma 
     \int_0^1 d\tau U_s(\kappa_1 \tilde x, z)
      [ \gamma_5 (2\tau -1)
      F_{\rho\lambda}(z) \nonumber \\
& & \quad\quad\quad\quad\quad\quad\quad -\; i
     {\tilde F}_{\rho\lambda}(z)]
    \tilde x^\lambda
     U_s(z, \kappa_2 \tilde x) \psi(\kappa_2 \tilde x) \Big\}
      +  (\kappa_1  \leftrightarrow \kappa_2), \nonumber
\end{eqnarray}
with $z=[(\kappa_1 -\kappa_2)\tau + \kappa_2)]\tilde x   $.
In  renormalized  quantum  field theory the proof of this representation 
is more complicated (for a rigorous derivation see
\cite{AZ}). 
In the following we use the light-cone gauge $\tilde{x}A =0 $,
 so that $U_s(\kappa_1 \tilde x, \kappa_2 \tilde x) \equiv 1 $.
Obviously, expression (\ref{lc1}) contains two light-ray operators
 from which we start our consideration:
\begin{eqnarray}
\label{p1.6}
  O_\rho(\kappa_1,\kappa_2) & =&
    {\bar\psi}(\kappa_1 \tilde x) \gamma_{\rho}
    \gamma_5 \psi(\kappa_2 \tilde x) , \\
\label{p1.7}
  O_{[\rho \sigma]} (\kappa_1,\kappa_2 ) &  = &
   {g\over 4} {\bar\psi}(\kappa_1 \tilde x)\gamma_\sigma
     \int_0^1 d\tau
      \big[ \gamma_5 (2\tau -1)
      F_{\rho\lambda}(z)
-  i {\tilde F}_{\rho\lambda}(z)\big]
    \tilde x^\lambda \psi(\kappa_2 \tilde x) 
\nonumber \\
& & - \;(\rho \leftrightarrow
    \sigma).
\end{eqnarray}
The operator  $O_\rho(\kappa_1,\kappa_2)$ contains twist two as well as 
twist three contributions, whereas $O_{[\rho \sigma]} (\kappa_1,\kappa_2)$
contains contributions of twist 3 and higher (because of their 
coefficient functions the last operator is power suppressed).

\section{Choice of  Light-Ray Operators}
\setcounter{equation}{0}
\renewcommand{\theequation}{\arabic{section}.\arabic{equation}}

As  mentioned before, the operator (\ref{p1.6}) has mixed twist, so we
look  for a decomposition into its parts of definite twist. For the local
light-cone  operators  there  exists  a  well-known procedure: 
 The  tensor  structure  of  operators  with  definite dimension
decomposes  into  irreducible  representations  of the (formal) symmetry
group  $O(4)$.  For light-ray operators, however, one has to take into
account towers of such irreducible representations.
Contraction  with  the  light-cone  vector $\tilde{x}$ projects onto the
leading twist two piece,
\begin{eqnarray}
\label{p1.8}
O^{\rm tw 2}(\kappa_1,\kappa_2) = \tilde{x}^\rho O_\rho(\kappa_1,\kappa_2)=
{\bar\psi}(\kappa_1 \tilde x)\not\!\tilde{x} \gamma_5 \psi(\kappa_2 \tilde x).
\end{eqnarray} 
The usual local twist two operators with spin $n$ follow according to 
$(\partial/\partial\kappa \equiv  \tilde{x}\partial)$ 
\begin{eqnarray}
\label{p1.8b}
O^{\rm tw 2}_{n} = 
 {(i\tilde x\partial)^{n-1}}O^{\rm tw 2}(0,\kappa)_{|\kappa=0} =
\tilde{x}^{\mu_1}\dots\tilde{x}^{\mu_n}  O^{\rm tw 2}_{\mu_1\dots\mu_n},
\end{eqnarray} 
where  $O^{\rm tw 2}_{\mu_1\dots\mu_n}$ is symmetrised and traceless.

In  order  to construct the twist three part we represent the operator
(\ref{p1.6})  in  terms  of  local  operators, extract the twist three
part  of  them,  and  express  the  result  again in terms of nonlocal
operators. It turns out that
\begin{eqnarray}
\label{p1.9}
O^{\rm tw 3}_\rho(\kappa_1,\kappa_2) &=&
-i(\kappa_2-\kappa_1)\int_0^1 du\, u\tilde{O}_\rho(\kappa_1,\kappa_1 \bar{u}+
\kappa_2 u),
\\
\label{p1.9b}
\tilde{O}_\rho(\kappa_1,\kappa_2) &=& 
     \int_0^{1} du\, 
     \bar{\psi}(\kappa_1\tilde{x}) i
   \gamma_{ [\rho}\,\tilde{x}_{\sigma ]}
   D^{\sigma}(u,\kappa_1\tilde{x},\kappa_2\tilde{x})
     \gamma^5 \psi([\kappa_1\bar u+\kappa_2 u]\tilde{x}),
\end{eqnarray}
where we introduced the notation 
$D^\rho(u,\kappa_1\tilde{x},\kappa_2\tilde{x})=
 \partial^\rho_{\kappa_2 \tilde{x}} + 
 ig A^\rho([\kappa_1 \bar{u}+\kappa_2 u]\tilde{x})$ and $\bar u = 1 - u $.
Analogous  to  the  twist two case our result can easily be checked by
comparing it with the local operators of definite twist.

All  operators  with  the  same  or  lower  twist  can  be mixed under
renormalization.  Therefore,  we  also have to take into account
 the following twist 3 operators:
\begin{eqnarray}
\label{p1.13}
{^\pm\! S_\rho}(\kappa_1,\tau,\kappa_2)&=& ig
    \bar{\psi}(\kappa_1\tilde{x})\not\!\tilde{x}
     \left[
      i \tilde{F}_{\alpha\rho}(\tau\tilde{x}) \pm
      \gamma^5 F_{\alpha\rho}(\tau\tilde{x}) 
     \right]\tilde{x}^\alpha
    \psi(\kappa_2\tilde{x}).
\end{eqnarray}
These  operators are the building blocks of the operator (\ref{p1.7}),
which is contained in the light-cone expansion. For the local
case,  the  importance  of  the operators (\ref{p1.13}) has been first
observed  by Shuryak and Vainshtein in \cite{SHV}. They result from an
application  of  the  equations  of  motion  to  the local twist three
operators.

The same can be shown directly and more easily also for the nonlocal 
operators.
Applying the equations of motion to the operator $\tilde{O}_\rho$ 
and using the relation 
$A_{\nu}(u\tilde{x})-A_{\nu}(v\tilde{x})=
\int^u_v d\tau \tilde{x}^\mu F_{\mu\nu}(\tau\tilde{x})$ 
which is valid in the light-cone gauge we get
\begin{eqnarray}
\label{p1.15}
& &\tilde{O}_\rho(\kappa_1,\kappa_2)
    = {i \over 2}(\kappa_2 - \kappa_1)  \int_0^1 du\, 
    \Big(\Omega^{\rm EOM}_\rho(u) + \Omega_\rho^{\rm REM}(u)
    -2 M_\rho(\kappa_1, \kappa_1\bar u +\kappa_2 u)
\nonumber\\     
& &\qquad\qquad\qquad+\;u\; 
{^+\!S_\rho}(\kappa_1,\kappa_1\bar u+\kappa_2 u,\kappa_2) +
   \bar u\; {^-\!S_\rho}(\kappa_1,\kappa_1\bar u+\kappa_2u,\kappa_2)\Big),
\end{eqnarray}
where $\Omega^{\rm EOM}$ 
is  an  equation of motion operator, $\Omega^{\rm REM}$ contains 
residual trace terms (being proportional to $\tilde{x}_\rho$)
and  operators containing an overall derivative, and $M_\rho$ is the
 mass dependent operator defined by
\begin{eqnarray}
\label{moper}
M_\rho(\kappa_1,\kappa_2)= 
        m\; \bar{\psi}(\kappa_1)
     \sigma_{\alpha\rho}\tilde{x}^\alpha \gamma^5
     (\tilde{x}D)(\kappa_2\tilde{x}) \psi(\kappa_2\tilde{x}),
\quad \sigma_{\alpha\beta}={i\over 2} [\gamma_{\alpha},\gamma_{\beta}].
\end{eqnarray}
Corresponding to Eq.\ (\ref{p1.9}) also the operator $O_\rho^{\rm tw 3}$
is build up of the 
Shuryak-Vainshtein operators ${^{\pm}\!S}_\rho$ and the mass operator $M_\rho$.
Because of the following properties the  additional operators  
$\Omega^{\rm EOM}$ and $\Omega^{\rm REM}$ can be neglected: If the operator
$\Omega^{EOM}$ is sandwiched between physical states their matrix element 
vanishes. 
The trace terms in the remaining operators $\Omega^{REM}$  can be safely
neglected and the obtained overall derivative vanishes in the forward case.

\section{Evolution Kernel of the Light-Ray Operators}
\setcounter{equation}{0}
\renewcommand{\theequation}{\arabic{section}.\arabic{equation}}

Before we present our result for the forward evolution kernel of
the nonsinglet twist three light-ray  operators we will discuss shortly
the mixing 
properties of these operators.  If we take into account all possible
mixing partners up to trace terms, then we get the following set of
operators
\begin{eqnarray}
\label{p1.18}
\left\{ \tilde{O}_\rho,\quad  {^\pm\!S_\rho},\quad M_\rho,\quad
 \Omega^{\rm EOM}_\rho \right\}.
\end{eqnarray}
Also  the  equation of motion operator $\Omega^{\rm EOM}_\rho$ possesses
an  anomalous  dimension  and  will be mixed with the other operators.
However,   from   the  general  renormalization  properties  of  gauge
invariant  operators  it  is  to  be expected that the counter term of
$\Omega^{\rm   EOM}_\rho$   is  only  given  by  the  operator  itself
\cite{JOG}.  In  fact,  this  was  explicitly  shown in \cite{BKL} and
recently  in  \cite{KO}.  Thus,  the  anomalous  dimension  matrix  is
triangular.  Forming  physical  matrix elements the equation of motion
operators  as  well  as their anomalous dimensions drop out completely
from  the  renormalization group equation. Furthermore, because of the
constraint    (\ref{p1.15})    we    can   eliminate   the   operator
$\tilde{O}_\rho$  and are left with the operators ${^\pm\!S_\rho}$ and
$M_\rho$ only.

There exists an additional property of the operators ${^\pm\!S_\rho}$
which guarantees that  ${^+\!S_\rho}$ and ${^-\!S_\rho}$ do not mix under 
renormalization. The contraction with 
$\epsilon_{\sigma\alpha\beta\rho}\tilde{x}^\alpha \tilde{x}^{\star\beta},$
where $\tilde{x}^{\star}$ is a second light-cone vector with
  $\tilde{x}\tilde{x}^{\star}=1$, transforms the operators 
${^\pm\!S_\rho}$ into $\pm {^\pm\!\hat{S}_\sigma}=
\pm {^\pm\!S_\sigma}\gamma^5$ (here the vertices of 
${^\pm\!\hat{S}_\sigma}$ and ${^\pm\!S_\sigma}$ differ by a 
$\gamma^5$ matrix). Since the anomalous dimensions of 
${^\pm\!\hat{S}_\sigma}$ and ${^\pm\!S_\sigma}$ coincide the
operators ${^\pm\!S_\rho}$  are effectively eigenstates of the duality 
transformation and thus do not mix under renormalization.

Using FeynCalc and some own subroutines for doing the momentum integration
in light-cone gauge  we calculated
the  pole part of the contributing one-loop diagrams of the operators
${^\pm\!S_\rho}$ and $M_\rho$ in space time dimension $n=4-2\epsilon$
(the  same  set  of  diagrams  as  given  in  \cite{BB}). 
Applying $\mu {d\over d\mu}=-\epsilon g{\partial\over \partial g} + \cdots$ 
we get the  result:
\begin{eqnarray}
\label{fresult1}
&&\hspace{-1.7cm}\mu^2{d\over d\mu^2} {^+\!S^\rho}({\kappa_1},{\kappa_2})=
       {\alpha_s\over 4\pi} \int_0^1 dy \int_0^{1-y} dz 
\nonumber\\ 
&&\hspace{-1cm} 
 \Big\{ (2C_F-C_A)\Big[
    y\delta (z) {^+\!S^\rho}(-{\kappa_1}y,{\kappa_2}-{\kappa_1}y) -
     2z {^+\!S^\rho}(\kappa_1-\kappa_2(1-z),-{\kappa_2}y)
\nonumber\\
& &\hspace{3.6cm} 
     + K(y,z) {^+\!S^\rho}({\kappa_1}(1-y)+
{\kappa_2}y,{\kappa_2}(1-z)+{\kappa_1}z)  
     \Big]   
\nonumber\\
  & &\hspace{-1cm}
 + C_A\Big[ 
  \Big(2(1-z)+L(y,z)\Big) 
      {^+\!S^\rho}({\kappa_1}-{\kappa_2}z,{\kappa_2}y)+
        L(y,z){^+\!S^\rho}({\kappa_1}y,{\kappa_2} - {\kappa_1}z) \Big]
\nonumber\\
 & &\hspace{-1cm}
+  2 C_F\,{{\left( 1 - y \right) }^2}\,\delta (1-z)\,
    M^\rho({\kappa_2}z-{\kappa_1}y)\Big\}
\end{eqnarray}
\begin{eqnarray}
\label{fresult2}
&&\hspace{-1.7cm} \mu^2{d\over d\mu^2} {^-\!S^\rho}({\kappa_1},{\kappa_2}) =
       {\alpha_s\over 4\pi} \int_0^1 dy \int_0^{1-y} dz 
\nonumber\\
&&\hspace{-1cm} \Big\{ (2C_F-C_A) 
    \Big[
     y\,\delta (z) 
     {^-\!S^\rho}({\kappa_1}-{\kappa_2}y,-{\kappa_2}y)- 
     2z {^-\!S^\rho}(-\kappa_1 y,\kappa_2 -\kappa_1(1-z)) 
\nonumber\\
&&\hspace{3.6cm} 
  +   K(y,z) {^-\!S^\rho}(\kappa_1\left(1-y \right)  + {\kappa_2}y,
          {\kappa_2}(1-z)  + {\kappa_1}z) \Big] 
\nonumber\\
&&\hspace{-1cm} 
+  C_A\Big[  
 \Big(2(1-z)+ L(y,z) \Big) 
     {^-\!S^\rho}({\kappa_1}y,{\kappa_2}-{\kappa_1}z) + 
     L(y,z) {^-\!S^\rho}({\kappa_1} - {\kappa_2}z,{\kappa_2}y) \Big] 
\nonumber\\
&&  \hspace{-1cm} 
+ 2C_F{{\left(1-z \right) }^2}\,\delta(1-y)\,  
M^\rho({\kappa_2}z-{\kappa_1}y)\Big\},
\\
&&  \hspace{-1.7cm} \mbox{where}
\nonumber\\
&& K(y,z) = 
 \left[
  1 + \delta (z)\,{1-y\over y}  + \delta (y)\,{1-z\over z}
 \right]_+
\nonumber\\
&& L(y,z) =
 \left[
  \delta (1 - y - z)\,{y^2\over 1-y} +
 \delta (z)\,{y\over 1-y} \right]_+
 - {7\over 4} \delta(1-y)\delta(z),
\nonumber   
\end{eqnarray}
and
\begin{eqnarray}
\label{fresult3}
 \mu^2 { d\over d\mu^2} M^\rho({\kappa}) = 
 -4 C_F  {\alpha_s\over 4\pi} 
 \int_0^1 dy \left(1+y-\left[{1\over 1-y}\right]_+ \right) 
        M^\rho({\kappa}y).
\end{eqnarray}
To condense the notation we used 
${^\pm\!S^\rho}(\kappa_1,\kappa_2)={^\pm\!S^\rho}(\kappa_1,0,\kappa_2)$, 
$M^\rho(\kappa)=M^\rho(0,\kappa) $, and the standard plus-prescription
fulfilling $\int dy [...]_+ = 0$ and $\int dy dz [...]_+ = 0$, respectively;
$C_F=4/3$ and $C_A=3$ are the usual Casimir operators of ${\rm SU}_c(3)$.

Since  charge conjugation transforms ${^+\!S^\rho}(\kappa_1,\kappa_2)$
into  ${^-\!S^\rho}(\kappa_2,\kappa_1)$  also the evolution kernels in
Eq.\ (\ref{fresult1}) and Eq.\ (\ref{fresult2}) are related to each
other.  It  is  easy  to  see  that  our kernels satisfy this symmetry
condition.  Up  to  a different definition of ${^\pm\!S^\rho}$ (and one
small  misprint)  our  result coincides with that of Balitsky and
Braun \cite{BB} restricted to the forward case and $m=0$.

Finally, we want to add some technical remarks.
 
Our  operators  are  not  completely  traceless.
Therefore, terms  proportional  to  $\tilde x_\rho$ have been omitted. 
Furthermore,
the auxiliary pole of the gluon propagator is regularized by the
Leibbrandt-Mandelstam prescription \cite{LM}
\begin{eqnarray}
\label{p1.22}
 \frac{1}{k\tilde{x} }= \frac{k \tilde{x}^\star}
   {(k\tilde x)(k\tilde{x}^\star) +i\epsilon}
\end{eqnarray}
with the light-like vector $\tilde{x}^\star$. There are different
advantages  of  this  prescription  proved  in  one  loop order, e.g.,
consistency  of  tensor  integral  relations and the validity of power
counting \cite{LM,KOR}. The investigation of local anomalous dimension
shows  that  $\tilde{x}^\star$-dependent operators as well as special
nonlocal  operators  may  appear  \cite{BAS}. However, their anomalous
dimensions  decouple  from the anomalous dimensions of gauge invariant
operators  \cite{JOG}.  Also  in  our  calculation a
 $\tilde{x}^\star$-dependent  operator appears and, as expected,
 does not contribute to the physical sector.

Let   us   point  out  shortly,  that  in  our  calculation  different
prescriptions  of the auxiliar pole in the gluon propagator 
do  only  affect  terms  which  are proportional to two
$\delta$-functions.  In the final result this effect is compensated by
the   wave  function  renormalization.  As an example,  we  consider  a
triangular  diagram of the operator ${^\pm\!S^\rho}$ with two external
fermion lines and one external gluon line
\begin{eqnarray}
\label{p1.23}
 \int d^{4-2\epsilon} k {P(\gamma,k, p_1,p_2)\over
 ((p_1 +k)^2 -m^2)((p_2 +k)^2 -m^2) k^2}
 {e^{i((p_1+k)\tilde x \kappa_1 -(p_2+k)\tilde x \kappa_2)}\over
 k\tilde x}.
\end{eqnarray}
The exponential  can be rewritten as
\begin{eqnarray}
\label{p1.24}
 e^{i(p_1\tilde x)\kappa_1 - (p_2\tilde x)\kappa_2}
  \left\{
   {e^{i(\kappa_1 - \kappa_2)k\tilde{x}}-1 \over k\tilde x } 
  + {1 \over k\tilde x} 
 \right\} .
\end{eqnarray}
The first term in the curly bracket is analytic in $k\tilde x$ so that 
the typical problem of light-cone gauge occure only in the second term 
which, however, contributes to a trivial $\kappa $-structure only.

\section*{Acknowledgment}

We wish to thank J.~Bl\"umlein, V.~M.~Braun, E.~A.~Kuraev, L.~N.~Lipatov, 
and O.~V.~Teryaev for valuable discussions. D.~M. was financially supported by
Deutsche Forschungsgemeinschaft (DFG).

\end{document}